\documentclass[preprint,12pt]{elsarticle}

\usepackage{epsfig}

\usepackage{amssymb}

\usepackage{setspace} 

\journal{arXiv.org}

\begin{document}
\begin{spacing}{1.5}  

\begin{frontmatter}
\title{Wave Model and Wave Drift Caused by the Asymmetry of Crest and Trough}

\author{Jin-Liang Wang}

\address{Research Institute for ESMD method and Its Applications,\\
College of science, Qingdao University of Technology, Shandong, P.R. China, 266520.\\
 E-mail: wangjinliang0811@126.com}


\begin{abstract}
It follows from the review on classical wave models that the asymmetry of crest and trough is the direct cause for wave drift. Based on this, a new model of Lagrangian form is constructed. Relative to the Gerstner model, its improvement is reflected in the horizontal motion which includes an explicit drift term. On the one hand, the depth-decay factor for the new drift accords well with that of the particle's horizontal velocity. It is more rational than that of Stokes drift. On the other hand, the new formula needs no Taylor expansion as for Stokes drift and is applicable for the waves with big slopes. In addition, the new formula can also yield a more rational magnitude for the surface drift than that of Stokes.\\[-5mm]
 \end{abstract}


\end{frontmatter}




\section{Introduction}
\setcounter{equation}{0}
The drift caused by water wave was firstly studied by George Gabriel Stokes in 1847.
Without other convincing models, his approximate formula
based on small-amplitude wave known as ``Stokes drift'' is taken as a default one till now.
\emph{Is the wave drift caused by the asymmetry of crest and trough?}
If the answer is true, then not only the nonlinear
Stokes wave with finite amplitude but also the Gerstner wave with large amplitude
exists wave drift. Thus the doubt of Weber (2011) can be well answered. This question
stimulates us to reconsider the wave mechanism.
Our answer is yes and the remodeling process
leads to a new formula for the wave drift which differs from that of Stokes.

In order to understand the wave mechanism, there is a necessity for us
to review the wave studies. Historical speaking,
the study of water wave can be dated back to the year 1687 when Newton
did an experiment with U-tube and
got the result ``the frequency of deep-water waves must be proportional
to the inverse of the square root of the wave length''.
As reviewed by Craik (2004), the classical wave theories were mainly developed
by the scientists from France, Germany and Britain
in the eighteenth and early nineteenth centuries.
Among all of them, the representative works are given
by Airy (1845) for linear wave, Stokes (1847) for nonlinear wave,
 Gerstner (1802) for trochoid wave and Earnshaw (1847) for solitary wave.
 After that time, the progresses are
 under the existing framework and on the wave-breaking investigation (Banner, 1993),
 the wind-wave growing mechanism (Phillips, 1957; Miles, 1957; Janssen, 2009),
 the wave-spectrum construction (Phillips, 1977; Wen and Yu, 1984)
 together with its applications in numerical ocean-wave forecast (Cavaleri et al, 2007; Mitsuyasu, 2002).
 One can also refer to the special issue ``Ocean Surface Waves''
 of \emph{Ocean Modeling}, Vol.70 (2013) for the latest developments on these aspects,
 such as those given by
 Tolman et al (2013) and Tolman $\&$ Grumbine (2013) for wave model improvements;
Perrie et al (2013) for wave-wave interactions;
Fedele et al (2013) for space-time measurements of oceanic sea states
and Benetazzo et al (2013) for wave-current interactions.
As for the study which takes Stokes drift as a special topic, they are
Mesquita et al (1992), Feng and Wiggins (1995), Jansons and Lythe (1998),
Webb and Kemper (2011), Liu et al (2014) and Myrhaug et al (2014), etc.
But most of them are about the applications of existing formula which was written down by Stokes in 1847.
To make remodeling it needs a new approach.
 Therefore, the present article only concern the classical results, especially the aspect of wave drift,
 given by Airy, Stokes and Gerstner. As for the solitary wave on shallow water
 given by Earnshaw, it is beyond the topic of periodic wave in deep water and is omitted here.
 The default form of it is the so-called ``gravity wave'' on the ocean surface.

\section{Classical Wave Models and Related Drift Arguments}

As the problem concerned, the default model
should be the inviscid and incompressible Navier-Stokes equations.
 But the solving of these equations involves in determining
the upper surface boundary condition which is just the wave to look for (Stewart, 2005).
This nonlinear characteristic makes the problem insoluble in essence.
So, the classical results for surface waves are merely some kind of approximations
and the drift formulas only hold within certain limits.

\subsection{On the Linear Wave Model}

 The classical linear wave theory illustrated in nowadays textbooks, such as those by
 Andersen $\&$ Frigaard (2011) and Soloviev $\&$ Lukas (2006), mostly follow from that of Airy (1845).
 Here the Cartesian coordinate system is adopted and only the 2-dimensional case is concerned.
The origin is chosen at the equilibrium level (the average height for the crest and trough)
with $x$ and $z$ pointing to the propagating direction
and upward direction separately.

On the assumption that the amplitude $A$ is infinitely small
relative to the wave-length $\lambda$ (related to the wave-number $k$ by $\lambda=2\pi/k$), that is, the wave steepness satisfies $\varepsilon=Ak\ll 1$ and
the upper boundary can be almost seen as a fixed flat surface,
 there is a linear approximation for the problem.
At this time, the surface traveling wave can be conjectured in the simplest trigonometric form:
\begin{equation}
\xi(x,t)=A\cos(kx-\omega t),
\end{equation}
here $\omega$ and $t$ denote the frequency and the time separately. For the deep-water case with irrotational hypothesis on the flow,
 the solving of the simplified Navier-Stokes equations yields depth-dependent profiles
 for the wave and pressure:
 \begin{eqnarray}
 &&\eta(x,z,t)=Ae^{kz}\cos(kx-\omega t),\\
 &&P(x,z,t)=P_0+\rho g\left[Ae^{kz}\cos(kx-\omega t)-z \right]
 \end{eqnarray}
together with a dispersion relation $\omega^2=gk$.
Here $\rho, g$ and $P_0$ are the water density, gravitational acceleration
 and constant air pressure on the surface.
At this time, the horizontal and vertical velocities are
\begin{eqnarray}
\left\{\begin{array}{ll}
 u(x,z,t)=A\omega e^{kz}\cos(kx-\omega t),\\[2mm]
 w(x,z,t)=A\omega e^{kz}\sin(kx-\omega t).
 \end{array}
 \right.
 \end{eqnarray}

According to the web of Wikipedia (2014), the derivation process of
the Stokes drift is as follows:

\emph{Within the framework of linear theory, the motion distance is very short and the particle's Lagrangian location $(x,z)$ can be substituted by the fixed equilibrium $(a,c)$ in (2.4) which yields the approximations:}
\begin{eqnarray}
\left\{\begin{array}{ll}
 x=a+\int u(a,c,t) dt=a-A\omega e^{kc}\sin(ka-\omega t),\\[2mm]
 z=c+\int w(a,c,t) dt=c+A\omega e^{kc}\cos(ka-\omega t).
 \end{array}
 \right.
 \end{eqnarray}
Based on this together with Taylor expansion technique, the Stokes drift is then estimated by:
 \begin{eqnarray}
U_s&=&\overline{u(x,z,t)}-\overline{u(a,c,t)}\nonumber\\
&=&\overline{u(a,c,t)+(x-a)u_a +(z-c)u_c+\cdots}-\overline{u(a,c,t)}\nonumber\\
&\approx& \overline{(x-a)x_{at} +(z-c)z_{ct}}\nonumber\\
&=&\overline{Ae^{kc}\sin{\theta}\cdot\omega kA e^{kc}\sin{\theta}
+ Ae^{kc}\cos{\theta}\cdot\omega kA e^{kc}\cos{\theta}}\nonumber\\
&=&\omega kA^2 e^{2kc}=\varepsilon^2 C_p e^{2kc}
 \end{eqnarray}
with $\theta=ka-\omega t$. Here the upper bar and subscripts denote the average
and partial derivative calculations separately. $C_p=\omega/k$ is the
phase speed of the propagation.

From the above analysis we see the formula for Stokes drift only holds
for $\varepsilon\ll 1$ and the magnitude of it is about $\varepsilon^2 C_p$
at the surface. It is known that, an ideal periodic motion
with closed trajectory can not result in a net drift. The generation of Stokes drift
should ascribe to the substitution of $(x,z)$ with $(a,c)$.
Under the small-amplitude hypothesis it seems reasonable for the approximation.
Yet the wave form given by eqns.(2.5) can not maintain ideal periodic motion
as eqn.(2.2) anymore, and its crest and trough have already possessed asymmetric characteristic.
As argued by Matioc (2010), for a linear wave no particle's-trajectory is closed,
unless the free surface is flat. This implies the shortcoming of the linear model.

\subsection{On the Stokes Wave Model}
In case $\varepsilon$ is not infinitely small,
there is a finite-amplitude wave model owing to Stokes (1847).
Notice that $\varepsilon=0.44$ accords with the critical case near broken (Massel, 2007),
its application range should be $0<\varepsilon\leq 0.44$.
With the aid of asymptotic expansion technique, the Stokes wave at the surface can be expressed as:
 \begin{eqnarray}
 \xi(x,t)=A\cos{\theta}+\frac{1}{2}\varepsilon A\cos{2\theta}
 +\frac{3}{8}\varepsilon^2A\cos{3\theta}+\cdots
 \end{eqnarray}
 with $\theta=kx-\omega t$ and $\omega^2=(1+\varepsilon^2+1.25\varepsilon^4+\cdots)gk$
 (Soloviev and Lukas, 2006; Stewart, 2005).
 The corresponding pressure profile approximates that of linear wave in eqn.(2.3).

For this case, the horizontal and vertical velocities are also in the forms of eqns.(2.4).
But the substitution of $(x,z)$ with $(a,c)$ is not suitable anymore. At this time,
the estimation of particle's trajectory is done relative to its initial location $(x_0,z_0)$.
By adopting the substitutions $x=x_0+h$ and $z=z_0+s$
together with approximating the equations for $h$ and $s$
it results in a Stokes drift $U_s=\varepsilon^2 C_p e^{2kz_0}$ (Wen and Yu, 1984)
which is same as that of linear wave with $z_0=c$.

Relative to the linear wave, the Stokes wave looses the range of wave steepness
to $0<\varepsilon\leq 0.44$ and it accords well with the actual one which has
 sharp crests and flat troughs. Its asymmetric characteristic is very distinct.

\subsection{On the Gerstner Wave Model}

On the assumption that the particle's trajectory is a circle,
Gerstner (1802) found a rotational trochoid wave:
\begin{eqnarray}
\left\{\begin{array}{ll}
x(a, c, t)=a-Ae^{kc}\cos(ka-\omega t),\\[2mm]
z(a, c, t)=c-Ae^{kc}\sin(ka-\omega t)
\end{array}
\right.
\end{eqnarray}
with a dispersion relation
$\omega^2=gk$. It is an exact solution of the two-dimensional Lagrangian equations
(Soloviev and Lukas, 2006):
\begin{eqnarray}
\left\{\begin{array}{ll}
x_{tt}x_a+z_{tt}z_a=-(P/\rho+gz)_a,\\[2mm]
x_{tt}x_c+z_{tt}z_c=-(P/\rho+gz)_c.
\end{array}
\right.
\end{eqnarray}

For this case, the water pressure is in a particular form (Wen and Yu, 1984; Weber, 2011):
\begin{eqnarray}
P&=&P_0-\rho g c-\frac{1}{2}\rho g A\varepsilon\left(1-e^{2kc} \right)\nonumber\\
&=&P_0+\rho g\left[-Ae^{kc}\sin(ka-\omega t)-z\right]\nonumber\\
& &-\frac{1}{2}\rho g A\varepsilon\left(1-e^{2kc} \right)
\end{eqnarray}
which has noting to do with the variables $a$ and $t$.
Here the last term reflect the effect from the fact that the equilibrium
 is higher than the motionless water level due to
the asymmetry of crest and trough.
This shows the water pressure is merely in the depth-dependent form $P(c)$
provided that the equilibrium $(a,c)$ is chosen as the reference frame.
In fact, to support the $t$-periodic wave motion the pressure should also vary
in a $t$-periodic manner. In this sense, the reference frame adopted here
has defect in describing the particle's motion, particularly for the drift
characteristic. A better choice for the reformation is to take the initial position
$(x_0,z_0)$ as a reference.

We note that the Gerstner model (2.8) is actually an alternative form
of the approximate linear model (2.5) with a translation on the phase angle by $-\pi/2$.
So the deduction process in eqn.(2.6) also holds for small $\varepsilon$.
This indicates the wave drift still exists for Gerstner model from the viewpoint of Taylor expansion.
Weber (2011) had ever doubted the net drift observed in wave tank experiments,
after all, the particles's trajectories of a Gerstner wave should be circles.
He had improved the model by adopting the viscosity.
However, the effect of viscosity to the gravity wave
is very small, its contribution to the wave drift should be limited.
There should be other deep reasons for this.

In addition, it follows from eqn.(2.8) that the particle's horizontal velocity at the wave crest equals to $u_c=A \omega$.
Notice that the wave breaks for $u_c>C_p=\omega/k$ (Massel, 2007),
the application range of wave steepness for Gerstner model should be
$0<\varepsilon\leq 1$.
Relative to the Stokes wave, its advantages lie in the concise expression and
the abandon of irrotational hypothesis.
To some extent, it accords better with the actual one which has sharp crests and flat troughs.
Along with the increasing of $\varepsilon$, the asymmetry of crest and trough
becomes serious and for big $\varepsilon$ the Taylor expansion around the
equilibrium may
result in big error which threaten the feasibility of Stokes drift formula.
Hence, there is a necessity for us to remodel the wave drift, particularly in the range
$0.44\leq \varepsilon\leq 1$.

In addition, it is easy to check that
\begin{eqnarray}
\left\{\begin{array}{ll}
x=Ut+a-Ae^{kc}\cos(ka-\omega t),\\[2mm]
z=c-Ae^{kc}\sin(ka-\omega t)
\end{array}
\right.
\end{eqnarray}
is also an exact solution to the Lagrangian equations in case a steady flow $U$ exists.
However, it follows from Clamond (2007)
 that the substitution of steady flow with Stokes drift
$U_s(c)=\varepsilon^2 C_p e^{2kc}$ is not permitted since
no steady wave exists of this form.

\section{Remodeling the Wave Motion}

\setcounter{equation}{0}
From the previous analysis we know Airy, Stokes and Gerstner
adopted a same approach, that is,
to take the conjectured wave forms as the preconditions.
What is more, the water pressures are given as corollaries in the last.
Here we take an inverse approach to do so.
\emph{Let the wave model be the object, the conjecture is done on the pressure.}

Take one water particle as the research object, we describe it
by Lagrangian coordinates $(x,z)$ with the initial
position $(x_0,z_0)$ as the reference.
We assume that the small particle possesses a cubic shape and it
maintains unchanged during the moving process.
Then it follows from Price (2006) that $\partial x/{\partial x_0}=\partial z/{\partial z_0}=1$
and $\partial x/{\partial z_0}=\partial z/{\partial x_0}=0$.
At this time, the eqns. (2.9) is simplified to
 \begin{eqnarray}
\frac{\partial^2 x}{\partial t^2}=-\frac{1}{\rho}\frac{\partial P}{\partial x},\qquad
\frac{\partial^2 z}{\partial t^2}=-\frac{1}{\rho}\frac{\partial P}{\partial z}-g.
\end{eqnarray}

\subsection{On the Pressure}

For a hydrostatic case with constant density, the water pressure increases linearly along with
the water-layer thickness $s$, that is, $P=P_0+\rho g s$.
When the fluid has a moving upper surface $z=\xi(x,t)$ it may also obey this rule with
\begin{equation}
P\approx P_0+\rho g [\xi(x,t)-z],
\end{equation}
this is the so-called ``quasi-hydrostatic approximation''
adopted in physical oceanography (Stewart, 2005). As the problem concerned,
if this kind of approximation is adopted,
then it follows from eqn.(3.1) that the vertical acceleration
${\partial^2 z}/{\partial t^2}\approx 0$. This means the vertical velocity almost keep unchanged.
It is impossible! The common sense is that the vertical velocities at the crest and the trough are all zero
but those at the mean level are not zero.

There is another case, might as well, call it by
``gravitational approximation'' which takes the gravity as the main restoring force.
For this case, there should be ${\partial^2 z}/{\partial t^2}\approx -g$ as the particle is on the upper crest part. For this case, $\partial P/{\partial z}\approx 0$.
This means there is no relative vertical force between two arbitrary water layers.
Hence, the horizontal pressure gradient force due to the slant water body is empty which leads to
${\partial^2 x}/{\partial t^2}\approx 0$.
This is also a strange case.

In fact, the quasi-hydrostatic and gravitational approximations are two extreme cases:
the vertical pressure gradient force is too strong for the first case and too weak for the second case.
Notice that the pressure formulas (2.3) and (2.10) for the linear, Stokes and Gerstner waves are deduced from the
 Navier-Stokes equations and their forms are very objective,
we follow them and estimate the pressure by
 \begin{eqnarray}
P=P_0+\rho g\left[e^{kz_0}\xi(x,t)-z \right].
\end{eqnarray}
Here the preconditioned sine or cosine function is substituted by an
undetermined free surface $z=\xi(x,t)$.
The modification is also reflected in the exponent,
use $kz_0$ to substitute $kz$ as in eqn.(2.10),
which accords well with the dynamic boundary condition $P=P_0$ at the surface for the case
$z_0=0$.
We note that the incorporating of $z_0$ here is permitted.
In fact, under the Lagrangian frame, the functions $x, z$ and $P$ can be all expressed by the variables $x_0, z_0$ and $t$.
 Yet, under the Euler frame whose variables are $x, z$ and $t$, it is strange to incorporate $z_0$ into eqn.(2.3).
As for the effect caused by the height difference between the equilibrium
and motionless water level, one can recall it back to improve the model.

\subsection{Model Construction}

To insert the pressure expression (3.3) into eqns.(3.1) it yields
 \begin{eqnarray}
\frac{\partial^2 x}{\partial t^2}=-g e^{kz_0} \frac{\partial \xi}{\partial x},\qquad
\frac{\partial^2 z}{\partial t^2}=-g k e^{kz_0} \xi.
\end{eqnarray}
Notice that the wave is a synthesis of transversal and longitudinal waves,
with the aid of these two equations we model them separately. To denote
 \begin{eqnarray}
\left\{\begin{array}{ll}
x(x_0,z_0,t)=x_0+e^{kz_0}X(x_0,0,t),\\[2mm]
z(x_0,z_0,t)=z_0+e^{kz_0}Z(x_0,0,t),
\end{array}
\right.
\end{eqnarray}
 then $x(x_0,0,t)=x_0+X(x_0,0,t),\; z(x_0,0,t)=Z(x_0,0,t)$ with $X(x_0,0,0)=Z(x_0,0,0)=0$.
At this time, the free surface possesses a Lagrangian description $(x_0+X, Z)$
 and an Euler description $Z=\xi(x_0+X,t)$.
With the new denotations eqs.(3.4) can be further simplified to
 \begin{eqnarray}
\frac{\partial^2 X}{\partial t^2}=-g\frac{\partial \xi}{\partial X},\qquad
\frac{\partial^2 Z}{\partial t^2}=-g kZ.
\end{eqnarray}
\emph{These mean the horizontal motion is due to the
pressure-gradient force caused by the slant water body
and the vertical motion is due to the variation of the surface elevation
itself (can be understood as the variation in the previous period,
it squeezes the water body and leads to new vertical motion).}

\subsubsection{Vertical Motion}

Before deriving the model of traveling-wave form we take no account of $x_0$
and only consider the motion of a surface particle begin from the point
$(0,0)$. The vertical component of it is determined by the second equation in (3.6).
Its solution reads:
   \begin{eqnarray}
Z=C_1\cos{\omega t}+C_2\sin{\omega t}
\end{eqnarray}
here $\omega^2=gk$, $C_1$ and $C_2$ are arbitrary constants.
In addition to the request $Z=0$ for the case $t=0$, might as well, we can limit it by
$Z=A$ for the case $t=\pi/{2\omega}$. To satisfy these two conditions,
we get an expression for the vertical motion:
\begin{eqnarray}
Z=A \sin{\omega t}.
\end{eqnarray}
It accords well with our common sense.

\subsubsection{Horizontal Motion}

The horizontal motion of the surface particle is determined by the first equation in (3.6).
It is associated with partial derivative of the undetermined
surface wave which is insoluble in essence. In the following we
estimate its solution by approximating the wave slope $\partial\xi/{\partial X}$.

Let $\delta$ be the average absolute value of $\partial\xi/{\partial X}$ over a wave-length $\lambda$ respect to the moment $t=0$, that is,
\begin{eqnarray}
 \delta&=&\frac{1}{\lambda}\int_0^{\lambda}\left|\frac{\partial\xi}{\partial X}\right|dX
\approx\frac{4}{\lambda}\left|\int_0^{\lambda/4}\frac{\partial\xi}{\partial X}dX\right|\nonumber\\
&=&\frac{4}{\lambda}|\xi(x_0+\lambda/4,0)-\xi(x_0,0)|=\frac{4}{\lambda}|-A-0|=\frac{4A}{\lambda},
 \end{eqnarray}
 here the position of wave trough is set on $x=x_0+\lambda/4$.
 We note that the commonly used wave steepness $\varepsilon=Ak$ is actually the maximum wave
 slope which relates to the mean one by $\varepsilon=\pi \delta/2$.
 Notice that the actual water wave has sharp crests and flat troughs, two other
parameters $\delta_1$ and $\delta_2$ are also borrowed to stand for the average wave slopes
on the crest and trough parts separately.

Notice that the vertical motion $Z=A \sin{\omega t}$ begins with
a rising process we approximate
the wave slope $\partial\xi/{\partial X}$ by two stags. For $0\leq t\leq T/2$,
the particle is on the upper crest part. In the first half time the crest is to the left
and the wave slope possesses the minimum value, say $-\varepsilon_1$, at $t=0$ and
$0$ at $t=T/4$. In the second half time the contrary is the case and the maximum value
$\varepsilon_1$ occurs at $t=T/2$. Also notice that the horizontal motion should
keep in step with the vertical one and follow the same change frequency,
we write it in the form $\partial\xi/{\partial X}\approx -\varepsilon_1 \cos{\omega t}$.
For $T/2< t\leq T$, the particle is on the lower trough part. The same deduction
process yields an approximate $\partial\xi/{\partial X}\approx -\varepsilon_2 \cos{\omega t}$.
Hence,
\begin{eqnarray}
\frac{\partial\xi}{\partial X}\approx\left\{
\begin{array}{ll}
 -\varepsilon_1 \cos{\omega t},\quad 0\leq t\leq T/2,\\[2mm]
-\varepsilon_2 \cos{\omega t},\quad T/2< t\leq T.
\end{array}
\right.
\end{eqnarray}
To inset this into the first equation of (3.6) it yields an estimation below:
\begin{eqnarray}
X=\left\{
\begin{array}{ll}
 A_1-A_1 \cos{\omega t},\quad 0\leq t\leq T/2,\\[2mm]
(2A_1-A_2)-A_2 \cos{\omega t},\quad T/2\leq t\leq T,
\end{array}
\right.
\end{eqnarray}
where $A_1$ and $A_2$ are the amplitudes of
horizontal motion relative to the crest and trough parts separately
which satisfy
\begin{eqnarray}
  A_1=\frac{\varepsilon_1}{k}=\frac{\varepsilon_1}{\varepsilon}A=\frac{\delta_1}{\delta}A,\quad
  A_2=\frac{\varepsilon_2}{k}=\frac{\varepsilon_2}{\varepsilon}A=\frac{\delta_2}{\delta}A.
  \end{eqnarray}
In addition, there is an interesting phenomenon that
$X=0$ for $t=0$ and $X=2(A_1-A_2)$ for $t=T$ and after a period of time
the particle propagates forward with a length $2(A_1-A_2)$ which implies a wave drift.

The remainder work is to find the relations between $\delta_1$, $\delta_2$ and $\delta$.
Since the asymmetry of crest and trough roots in the horizontal motion
and their difference ascribes to the last stage, there should be
  \begin{eqnarray}
\frac{\lambda}{4}+A_2=\frac{A}{\delta_2},\quad \frac{\lambda}{4}-A_2=\frac{A}{\delta_1}.
\end{eqnarray}
It follows from eqns.(3.12) and (3.13) together with the relationship ${\lambda}/{4}={A}/{\delta}$ that
 \begin{eqnarray}
\delta_1=\frac{3+\sqrt{1+4\delta}}{4-2\delta}\delta,\quad \delta_2=\frac{\sqrt{1+4\delta}-1}{2}.
\end{eqnarray}
Their variations are depicted in \emph{Figure 1}.
On the one hand, it shows that the ratios $\delta_1/\delta, \delta_2/\delta \longrightarrow 1$ as $\delta\longrightarrow 0$.
This means \emph{the smaller the average wave slope the better the symmetry for the crest and trough}.
In case the slope becomes small enough, the wave surface can be approximated by the linear model.
On the other hand, it shows that the crest slope $\delta_1$ increases and the trough slope $\delta_2$ decreases
relative to the average one $\delta$ as it increases. This means \emph{the bigger the average wave slope
the sharper the crest and in case the slope becomes big enough the wave may firstly break at the top of the crest}.

\begin{figure}
  \centerline{\includegraphics[height=2.8in,width=4in]{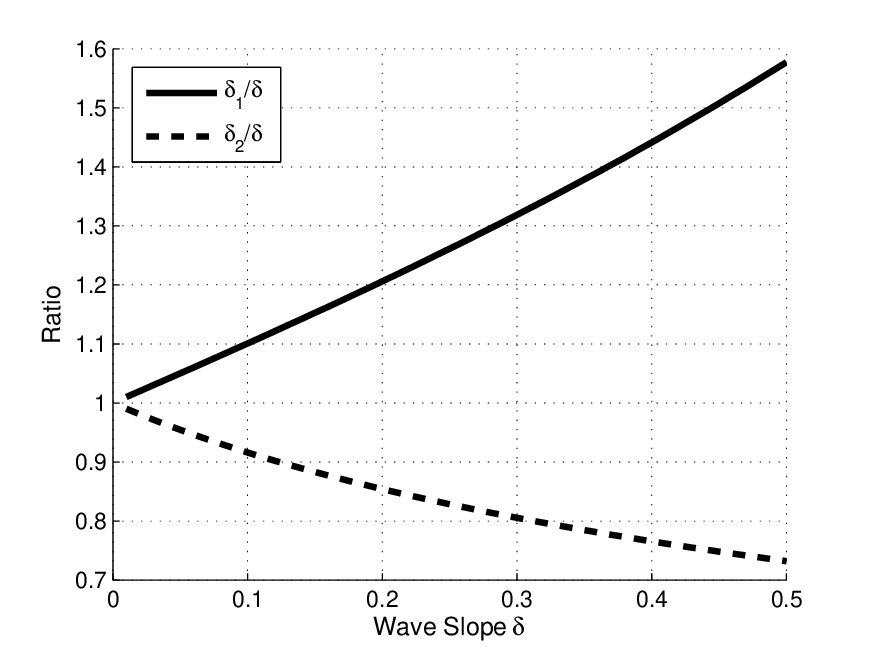}}
  \caption{\footnotesize The relative variation of the slopes $\delta_1$ and $\delta_2$ for the crest and
  trough parts along the average one $\delta$. }
\end{figure}

\subsubsection{Model in Traveling Wave Form}

Now that the particle's horizontal and vertical motions are constructed,
it is time for us to recall back the transformation (3.5).
Since the equilibrium $(a,c)$ is more convenient than the initial position
$(x_0,z_0)$ in describing traveling wave, here we return to the common way
with the transforms $a=x_0+A_1 e^{kz_0}$ and $c=z_0$. Therefore, the horizontal motion
reads
\begin{eqnarray}
x=a+e^{kc}\left\{
\begin{array}{ll}
 -A_1 \cos{\omega t},\quad t\in [0, T/2],\\[2mm]
(A_1-A_2)-A_2 \cos{\omega t},\quad t\in [T/2,T].
\end{array}
\right.
\end{eqnarray}
To substitute $\omega t$ by $\theta=ka-\omega t-2n\pi$ it leads to a
traveling-wave form:
\begin{eqnarray}
x=a+n\alpha  e^{kc}+e^{kc}\left\{
\begin{array}{ll}
 -A_1 \cos{\theta},\quad \theta\in [0, \pi],\\[2mm]
\alpha/2-A_2 \cos{\theta},\quad \theta\in [\pi, 2\pi],
\end{array}
\right.
\end{eqnarray}
where $\alpha=2(A_1-A_2)$, $n=0,1,2,\cdots$. The similar deduction process for the vertical motion
yields
\begin{eqnarray}
z=c-A e^{kc}\sin{\theta}.
\end{eqnarray}
The corresponding dispersion relation still remains $\omega^2=gk$.

The above two equations compose a new water wave model.
It differs from the linear model, nonlinear Stokes model and Gerstner model.
From \emph{Figure 2} we see the newly derived model and Gerstner model
are better than the Stokes one in reflecting the crest-trough asymmetric characteristic.
Relative to the Gerstner model in (2.8), the improvement of the new one lies in the horizontal component
which includes an explicit drift term.
In fact, it follows from the modeling process that
\emph{the wave drift is mainly caused by the asymmetry of crest and trough.}
The Stokes drift for the linear model and Stokes model is merely
an indirect reflection to this point.

\begin{figure}
  \centerline{\includegraphics[height=2.8in,width=4in]{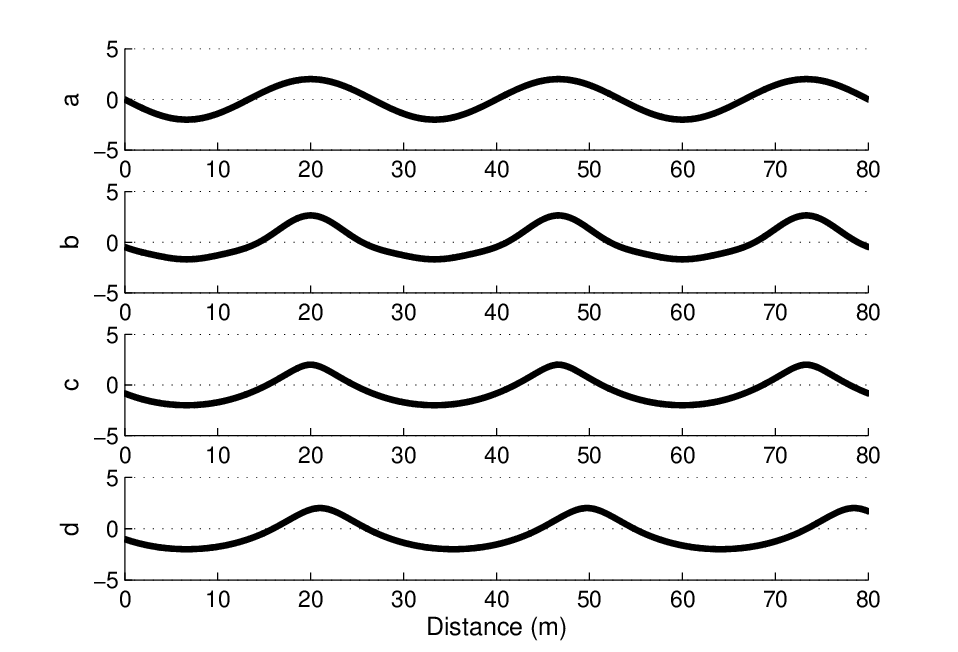}}
  \caption{\footnotesize Comparison among four wave models for $A=2$m and $\delta=0.3$. \textsf{a}: the linear model;
\textsf{b}: the third-order Stokes model; \textsf{c}: the Gerstner model and \textsf{d}: the newly derived model. }
\end{figure}

\section{New Wave Drift Formula}

It follows from eqn.(3.16) that, on each period of time $T$ all the particles propagate forward with
the same length $\alpha  e^{kc}$ (see \emph{Figure 3}). So there is an average velocity for the wave drift:
 \begin{eqnarray}
U_d&=&\frac{\alpha e^{kc}}{T}=\frac{2(\delta_1-\delta_2)A}{\delta T} e^{kc}
=\frac{\delta_1-\delta_2}{2}C_p e^{kc}\nonumber\\
&=&\frac{1+\delta-(1-\delta)\sqrt{1+4\delta}}{2-\delta}\sqrt{\frac{gA}{2\pi\delta}} e^{kc}
\end{eqnarray}
for $0<\delta <2$, here the relations in eqns.(3.12) and (3.14)
together with the transform $C_p=\omega/k=\sqrt{g/k}=\sqrt{2Ag/{\pi\delta}}$ are used.
It is easy to see the wave drift depends not only on the wave amplitude $A$, but also
on the wave slope $\delta$ and water depth $c$.

\begin{figure}
  \centerline{\includegraphics[height=2.8in,width=4in]{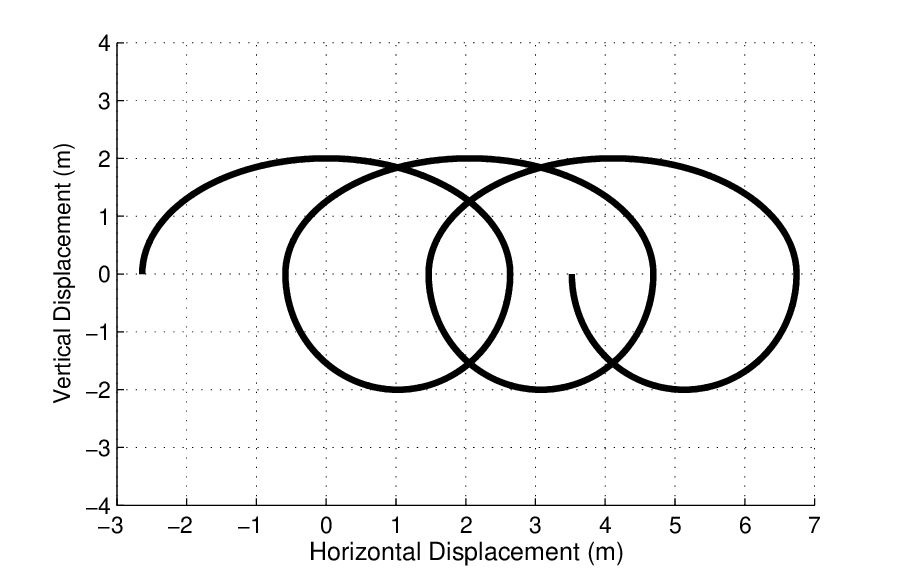}}
  \caption{\footnotesize The surface-particle's trajectory
  respect to the newly derived model with amplitude $A=2$m and slope $\delta=0.3$. }
\end{figure}

Relative to the Stokes drift
 \begin{eqnarray}
U_s=\varepsilon^2 C_p e^{2kc}=\sqrt{\frac{g\pi^3\delta^3A}{8}} e^{2kc}
\end{eqnarray}
the modifications of new formula are reflected in the depth-decay and
slope-dependent factors. \emph{Since the horizontal velocity of the water particle
has a depth-decay factor $e^{kc}$, it is natural for the wave drift
possessing the same one.} On the contrary, the factor $e^{2kc}$ seems strange.
As for the slope-dependent factor, the estimation of Stokes drift is done by
Taylor expansion around the particle's equilibrium which requires
a small wave slope $\delta$. Though the application of it is extended from the
linear model to the nonlinear Stokes model, its applicable scope still remains
$0<\delta\leq 0.44\cdot 2/\pi\approx 0.28$. \emph{Yet the new formula
is directly modeled from the wave mechanism, it needs no
expansion management and the applicable scope is extended to $0<\delta \leq 0.43$.} Here
the upper bound is an approximation to the limiting case near broken.
In fact, it follows from (3.11) that the surface particle possesses a maximum horizontal
velocity $u_c=A_1\omega$ at the crest. \emph{A breaking wave requires $u_c> C_p=\omega/k$
which yields a breaking criterion $\delta_1>2/\pi$, that is,
the wave breaks when the average slope angle of the crest
is bigger than $\arctan(2/\pi)=32.48^{\circ}$ which is a little bigger than the known breaking
criterion $30^{\circ}$ (Massel, 2007).} In term of the commonly used wave steepness,
it accords with the critical value $\varepsilon_1=1$
which corresponds to the steepest angle $45^{\circ}$ at the down-most of the crest.
As for the critical value $\delta=0.43$,
it is solved from the equation $\delta_1= 2/\pi$ with referring to the relationship
between $\delta_1$ and $\delta$ in eqn.(3.14).

In the following we compare the newly derived formula with that of Stokes drift by numerical approach. Here only the surface drift is considered.
It follows from \emph{Figure 4} that
the newly derived formula yields a surface drift $0.45$m/s
whose magnitude is more rational than that of Stokes ($1.29$m/s) at its upper applicable bound $\delta=0.28$.
Even at the renewed upper bound $\delta= 0.43$, the one given by the new formula is not bigger
than $0.84$m/s, yet that of Stokes attains $2.46$m/s which is too strong to meet
 the common sense.

 \begin{figure}
  \centerline{\includegraphics[height=2.8in,width=4in]{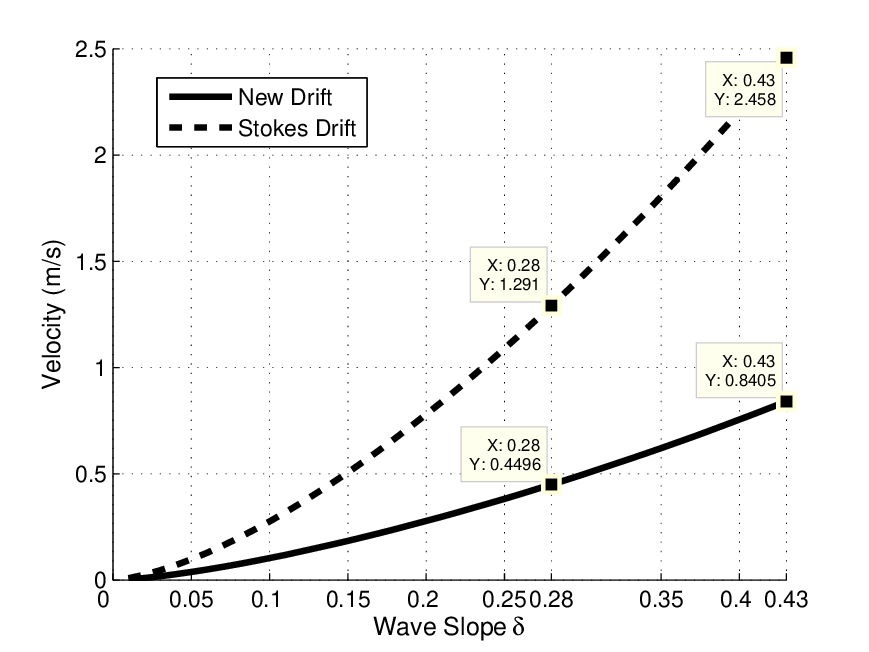}}
  \caption{\footnotesize Comparison between the newly derived wave drift and Stokes drift
  respect to a surface wave with amplitude $A=2$m along with the variation of wave slope $\delta$. }
\end{figure}

\section{Conclusions}

By reviewing the classical linear wave, Stokes wave and Gerstner wave
we have found that \emph{the asymmetry of crest and trough is the direct cause for wave drift.}
Based on this, a new model of Lagrangian form is constructed.
Relative to the Gerstner model, its improvement is reflected in the horizontal component
which includes an explicit drift term.
The newly derived drift formula depends not only on the wave amplitude $A$, but also
on the average wave slope $\delta$ and water depth $c$.

On the one hand, the depth-decay factor $e^{kc}$ for the new drift accords well with
that of the particle's horizontal velocity. It is more rational than
$e^{2kc}$ in Stokes drift.
 On the other hand, the estimation of Stokes drift is done by
Taylor expansion around the particle's equilibrium which requires
an applicable scope $0<\delta\leq 0.28$. Yet the new formula
is directly modeled from the wave mechanism, it needs no
expansion management and the applicable scope is extended to $0<\delta \leq 0.43$.

To estimate the drift of big waves at sea are valuable for ocean engineering.
 A good formula should be able to yield a reliable magnitude for it.
The numerical simulations show that the newly derived formula yields a more rational surface drift ($0.45$m/s$\leq U_d\leq 0.84$m/s)
 than that of Stokes one ($1.29$m/s$\leq U_s\leq 2.46$m/s) for the case $0.28\leq \delta\leq 0.43$.
In fact, it is rare to observe a current with velocity of several knots at sea, not to say the drift
 of particle's trajectory.

\noindent\textbf{Acknowledgments.} We thank the supports from
 the National Natural Science Fund of China (No.41376030).

\end{spacing}

\end{document}